\documentclass[12pt]{amsart}
\usepackage{amssymb,amscd}
\newcommand\ts{\mkern 2mu}
\newcommand{\bx}{\boldsymbol{\times}}
\newcommand{\br}{\mathbf r}
 \newcommand{\ri}{{\mathrm i}\mkern1mu}
\newcommand{\rd}{{\mathrm d}\mkern1mu}
\DeclareMathOperator*{\GL }{{\mathsf {GL}}}
\DeclareMathOperator*{\SL }{{\mathsf {SL}}}
\DeclareMathOperator*{\SO }{{\mathsf {SO}}}
\DeclareMathOperator*{\Ort }{{\mathsf {O}}}
\DeclareMathOperator*{\Spin }{{\mathsf {Spin}}}
 \DeclareMathOperator*{\Pin}{{\mathsf {Pin}}}
 \DeclareMathOperator*{\Cl}{{\mathsf{Cliff}}}
 \DeclareMathOperator*{\End }{{\mathsf {End}}\ts}
 \DeclareMathOperator*{\id}{{\mathrm id}}
\DeclareMathOperator*{\Ad}{{\mathrm Ad}}

\begin{document}

\vspace*{-2.5cm}

\begin{center}{{\small Based on the text of a talk given at the 
Conference}\\{\small \sl Particles, Fields, and Gravitation `98\/}
{\small (\L\'od\'z, 14--18 May 1998)}\\{\small Proceedings 
to be published by the 
American Institute of Physics}}
\end{center}
\vspace{1cm}

\title{Reflections and spinors on manifolds}
\date{}

\author{Andrzej Trautman}

\address{
Instytut Fizyki Teoretycznej, Uniwersytet Warszawski\\
Ho\.za 69, 00-681 Warszawa, Poland}
\email{amt@fuw.edu.pl}
\thanks{Research supported in part 
by  the Polish Committee for Scientific Research (KBN) under grant 
no. 2 P03B 017 12 and by 
 the Foundation for Polish-German
Cooperation with funds 
 provided by the Federal Republic of Germany.\\
\hspace*{1em}\( \! \)  This paper is dedicated to 
the memory of Ryszard R\c aczka.}

\begin{abstract}
This paper reviews some recent work on (s)pin  structures and the Dirac 
operator on hypersurfaces  (in particular, on spheres), on 
real projective spaces and quadrics. Two 
approaches to spinor fields on manifolds 
are compared. The action of reflections 
on spinors is discussed, also for two-component (chiral) spinors. 
\end{abstract}

\maketitle
\section*{Introduction}
This paper contains a brief review of the work, done mostly in 
collaboration with Ludwik D\c abrowski \cite{DT}, 
Michel Cahen, Simone Gutt \cite{CGT} and Thomas 
Friedrich \cite{FT}, on pin structures and the Dirac operator on 
higher-dimensional Riemannian manifolds; see  also \cite{AT2} 
and the references given there. In physics, there is now interest in 
higher dimensions motivated by research on unified theories, on 
supersymmetries, strings and their generalizations. There is also an 
intrinsic motivation: the Dirac operator is a fundamental object of 
an importance comparable to that of the Laplace operator and of the 
Maxwell, Yang-Mills and Einstein equations. 

Reflections in a quadratic space generate the orthogonal group of 
automorphisms of that space; according to the Cartan-Dieudonn\'e 
theorem, every orthogonal 
transformation in an \( m \)-dimensional quadratic space can be 
written as the product of a sequence of no more than \( m \) reflections.
 Reflections are of considerable interest in physics: 
invariance of electromagnetism 
under space reflections leads to selection rules; their 
violation is a striking feature of weak interactions. The PCT theorem 
describes a fundamental property of 
relativistic quantum field theories.

There are several ``spinorial'' extensions of orthogonal groups; each 
of them can be used to define  a ``(s)pin structure'' that is required to 
describe spinor fields on a curved manifold.  For these 
structures to  exist, the manifold should satisfy certain topological 
conditions; see \cite{F} and the literature listed there. In 
this paper, I recall two definitions of spinor fields on manifolds 
and give several simple examples of spin and pin 
structures. The next section summarizes the definitions and 
terminology of Clifford algebras and their representations.

\section*{Clifford algebras and spinors}
There are important differences---and similarities---between spinors 
associated with vector spaces of even and odd dimensions. 

Let \( h \) 
be a quadratic form on a real vector space \( V \) of 
dimension \( m \). The pair \( (V,h) \) is said to be a quadratic space. 
The Clifford algebra \( \Cl(h) \) associated with \( (V,h) \) is 
generated by elements of \( V \) subject to relations of the form
\( u^{2}=h(u) \); see \cite{BR,LD,F} and Ch. IX of  \cite{Bo}.
 Let \( \alpha \) be the involutive automorphism of 
\( \Cl(h) \) such that \( \alpha(1)=1 \) and \( \alpha(v)=-v \) for 
every \( v\in V \). This (main) automorphism defines
 a \( {\mathbb Z}_{2} \)-grading of the Clifford {\it algebra\/}, 
\( \Cl(h)=\Cl^{\rm even}(h)\oplus\Cl^{\rm odd}(h) \) 
and \( V \) is a 
vector subspace of the odd part. Let \( (e_{i}), i=1,\dots,m \)
be an orthonormal frame in \( V \). As a
 {\it vector\/}  space, the algebra \( \Cl(h) \) is \( {\mathbb Z} 
\)-graded, \( \Cl(h)=\oplus_{p=0}^{m}\Cl^{p}(h) \), where
 \( \Cl^{p}(h) \) is the vector space spanned by all
elements of the form \( e_{i_{1}}\dots e_{i_{p}} \) such that
\( 1\leqslant i_{1}<\dots <i_{p}\leqslant m \). In particular, 
\( \Cl^{m}(h) \) is spanned by the {\it volume element\/}
 \( \eta=e_{1}\dots e_{m} \); its 
square is either 1 or \( -1 \), depending on the signature of \( h 
\). If \( u\in V \) is not null, \( u^{2}\neq 0 \), then the linear map \( 
V\to V \), \( v\mapsto -uvu^{-1} \),
is a reflection in the hyperplane orthogonal to \( u \). Since \( 
\eta v=(-1)^{m+1}v\eta \) for every \( v\in V \), if \( m \) is {\it 
even}, 
then one  can  write 
\(-uvu^{-1}=(u\eta)v(u\eta)^{-1}  \). One says that \( u\in V \) is a unit 
vector if \( h(u)=1 \) or \( -1 \). The group \( \Pin(h) \) is defined as the 
set of products \( u_{1}u_{2}\dots u_{r} \) of all sequences of unit 
vectors, with a composition induced  by Clifford multiplication
and \( \Spin(h)=\Pin(h)\cap \Cl^{\rm even}(h) \).
The {\it adjoint\/} representation \( \Ad \) of \( \Pin(h) \) in \( V \)
is defined by \( \Ad(a)v=ava^{-1} \), where \( a\in\Pin(h) \) and \( 
v\in V \). For \( m \) {\it even\/}, 
the map \( \Ad \) is a homomorphism onto \(\Ort(h)  \)
 with kernel \( {\mathbb Z}_{2}=\{1,-1\} \); for \( m \) {\it odd\/}, 
\( \Ad \) is a homomorphism onto \( \SO(h) \). In both cases, 
 to obtain a double cover of \( \Ort(h) \) that coincides with \( \Ad \)
when restricted to \( \Spin(h) \), one can use the {\it twisted 
adjoint\/} representation \( \widetilde{\Ad} \) of \( \Pin(h) \), defined 
by \( \widetilde{\Ad}(a)v=\alpha(a)va^{-1} \). If  \( h \) is of signature
\( (k,l) \), \( k+l=m \), then one writes \( \Cl_{k,l} \), \( 
\Pin_{k,l} \), and \( \Spin_{k,l} \) instead of \( \Cl(h) \), \( 
\Pin(h) \) and \( \Spin(h) \), respectively.

For \( m=2n \), the algebra \( \Cl(h) \) is central simple and has 
one, up to equivalence, irreducible and faithful 
representation  \( \gamma \) in a
complex, \( 2^{n} \)-dimensional space \( S \) of {\it Dirac\/} spinors. 
Given such a representation, one identifies \( \Cl(h) \) with its 
image in \( \End S \).
Upon restriction to \( \Cl^{\rm even}(h) \), this representation 
decomposes into the direct sum of two irreducible and 
complex-inequivalent representations in 
spaces of Weyl (chiral, reduced or half) spinors so that \( 
S=S_{+}\oplus S_{-} \). The Dirac operator 
changes the chirality of spinors. Introducing the \( 2^{n}\times 2^{n} 
\) Dirac matrices \( \gamma^{i} \) and writing \( \gamma_{\pm}^{i}
=\gamma^{i}\mid
S_{\pm} \), one obtains the well-known decomposition of the Dirac 
operator, \( \gamma^{i}\partial_{i}=\bigl(\begin{smallmatrix} 0&
\gamma_{-}^{i}\partial_{i}\\\gamma_{+}^{i}\partial_{i}&0
\end{smallmatrix}\bigr)\).

For \( m=2n-1 \), the algebra \( \Cl^{\rm even}(h) \) is central 
simple and has one, up to equivalence, representation in a complex, 
\( 2^{n-1} \)-dimensional space of {\it Pauli\/} spinors. The full 
algebra has two complex-inequivalent, in general not faithful, 
representations in spaces of Pauli spinors. The direct sum of these 
representations is a decomposable, but faithful, representation of
\( \Cl(h) \) in the \( 2^{n} \)-dimensional space of {\it Cartan\/} 
spinors. Similarly as  in this case of an even-dimensional space,
 the Dirac operator interchanges here the 
spinors belonging to the two Pauli representations. Namely, 
let \( \sigma^{i} \), where \( i=1,\dots,2n-1 \),  
 be the \( 2^{n-1}\times 2^{n-1}  \) Pauli matrices. 
The (modified) Dirac operator, acting on 
Cartan spinor fields, can be written 
as \(\bigl(\begin{smallmatrix}0&\sigma^{i}\partial_{i}\\
\sigma^{i}\partial_{i}&0\end{smallmatrix}\bigr)\). The Cartan 
representation is essential when one considers the Dirac operator on 
non-orientable, odd-dimensional manifolds \cite{CGT1,AT1}.

\section*{Spinors on manifolds}
There are (at least) two approaches to spinors on manifolds; both of 
them can be traced  to early work by mathematicians and 
physicists; 
see \cite{Sch} and the references to the period 1928-1931 given there. 

\subsection*{The classical approach}
 The first
approach to be summarized here, initiated by Wigner, Weyl and Fock,
 consists in referring spinors to tetrads (``Vierbeine''); 
its modern formulation uses the notion of  a (s)pin structure 
involving a ``prolongation'' of the bundle \( P \) of orthonormal 
frames to the principal 
(s)pin bundle \( Q \). More precisely, given a Riemannian manifold \( M 
\) with a metric tensor of signature \( (k,l) \), a \( {\Pin}_{k,l} 
\)-structure on \( M \) is given by the maps
\begin{equation}
\begin{CD}
{\Pin}_{k,l}      @>>>Q\\
 @V\widetilde{\Ad} VV  @VV {\chi} V\\
 {\Ort}_{k,l}       @>>>P    @>\pi >>M,
 \end{CD}
\label{e:defpinstr}
\end{equation} 
such that \( \chi(qa)=\chi(q)\widetilde{\Ad}(a) \), 
\( (q,a)\mapsto qa \) denotes 
the action map of \( {\Pin_{k,l}} \) in the principal spin bundle \( Q 
\), etc. If \( {\Pin}_{k,l} \) in \eqref{e:defpinstr} is replaced by \( 
{\Pin}_{l,k} \), then one obtains the definition of a \({\Pin}_{l,k} 
\)-structure. They are both referred to as pin structures;
 if the manifold is orientable and has a pin structure, then it has a 
spin structure. The diagram describing a spin structure is shortened 
to
\begin{equation}
{\Spin}_{k,l}\to Q\to P\to M.
\label{e:sstr}
\end{equation}
The spinor connection is obtained as the lift 
of the Levi-Civita connection from \( P  \) to \( Q \). This approach, 
 standard in mathematics \cite{F},   
is sometimes criticized by physicists who 
say that they have no use for 
{\it principal bundles\/} and are willing to 
consider only spinor fields.

One can present this approach in a language familiar to physicists, 
by referring everything to local sections of the bundle \( Q\to M \) 
and using the terminology of {\it gauge fields}. 
For simplicity, consider an 
even-dimensional manifold, \( k+l=2n \), put \( G={\Pin}_{k,l} \) and 
let a representation of \( {\Cl}_{k,l} \) in  \( S \) be given by 
the Dirac matrices \(\gamma_{i}\in{\End} S  \). A spinor field is now 
a map \( \psi:M\to S \); given a function  \( U:M\to G \), one 
defines the gauge-transformed spinor field as \( \psi'=U^{-1}\psi \),
\( \psi'(x)=U(x)^{-1}\psi(x) \) for \( x\in M \). A spinor connection 
(``gauge potential'') is a 1-form \( \omega \) on
 \( M \) with values in the Lie algebra of
\( G \), i.e. in \( 
\Cl^{2}_{k,l}\subset \End S\); therefore, it can be written as \( 
\omega=\tfrac{1}{4}\gamma^{i}\gamma^{j}\omega_{ij} \), where \( 
\omega_{ij}=-\omega_{ji} \) are 1-forms. 
 The covariant (``gauge'') derivative  of \( \psi \) is 
\begin{equation}
 D\psi=\rd \psi +\omega\psi. 
\label{e:defD}
\end{equation}
The gauge transformation \( U \) induces a change of 
the connection, \( \omega\mapsto \omega'=
U^{-1}\omega U+U^{-1}\rd U \) 
so that \( (D\psi)'=U^{-1}D\psi \). Since the dimension of \( M \) is 
even,  the adjoint representation is onto \( {\Ort}_{k,l} \) and one 
can define, for every \( a\in{\Pin}_{k,l} \subset \GL(S)\),  the 
(orthogonal) matrix \( ({\rho^{i}}_{j}(a) )\) by
\(
a^{-1}\gamma^{i}a={\rho^{i}}_{j}(a)\gamma^{j}\),
so that
\(a^{-1}\gamma_{i}a=\gamma_{j}{\rho^{j}}_{i}(a^{-1})\).
From the Lemma:
if \( a\in{\Cl}^{p}_{k,l}\), then \(
g^{ij}\gamma_{i}a\gamma_{j}=(-1)^{p}(n-2p)a\),
taking into account that \( U^{-1}\rd U \) is in the Lie algebra of \( 
G \)---therefore of degree \( p=2 \)---one obtains
\(
g^{ij}U^{-1}\gamma_{i}U\rd(U^{-1}\gamma_{j}U)=4\ts U^{-1}\rd U
\)
so that 
\(
{{\omega'}^{i}}_{j}={\rho^{i}}_{k}(U^{-1}){\omega^{k}}_{l}
{\rho^{l}}_{j}(U)+
{\rho^{i}}_{k}(U^{-1})\rd {\rho^{k}}_{j}(U)
\).
Let \( (e_{i}) \) be a field of orthonormal frames on \( M \) and let
\( (e^{i}) \) denote the dual field of coframes.
Since, by definition, \( \omega_{ij}+\omega_{ji}=0 \), the 1-forms 
\( (\omega_{ij}) \) define a {\it metric\/} 
linear connection. Its {\it torsion\/}
\( \rd e^{i}+{\omega^{i}}_{j}\wedge e^{j} \) need not be zero. 

The action of the Dirac operator \( {\mathcal D} \) on a spinor field 
is  \( {\mathcal D}\psi=\gamma^{i}e_{i}\lrcorner D\psi \), where
\( \lrcorner  \) denotes contraction.

\subsection*{Spinor fields according to Schr\"odinger and Karrer}
 The  second 
 approach can be traced back to work by Tetrode; it has been clearly 
formulated by Schr\"odinger \cite{Sch}\footnote{I thank Engelbert Sch\"ucking 
for having drawn my attention to this remarkable paper. It
contains a derivation of the formula for the square of the 
Dirac operator on Riemannian manifolds.} and Karrer \cite{K}; it
sometimes appears in texts written by 
physicists; see, e.g., \cite{BT}. Consider 
a Riemannian manifold \( (M,g) \) and let \( g_{x} \)
 denotes the quadratic 
form induced by \( g \) in the 
vector space \( T_{x}M \) tangent to \( M \) at \( x \). 
 One assumes now the existence of a 
representation \( \gamma\)  of the Clifford 
bundle \( {\Cl}(g)=\bigcup_{x\in M}\Cl(g_{x}) \) 
 in a {\it vector  bundle\/} 
\( \varSigma\to M \) of  spinors so that \( 
\gamma(u)^{2}=g(u,u){\id}_{T_{x}M} \) for every  \( u\in T_{x}M \).
 One then  introduces a spinor covariant derivative  on \( \varSigma \), 
compatible with a metric covariant derivative on \( TM \). Such a 
structure is weaker (more general) than a classical spin structure. 
For example, it exists on every almost Hermitean manifold even 
though some of these manifolds---such 
as the even-dimensional complex 
projective spaces---do not admit a spin structure.
The precise relation between those two 
approaches, in the case of even-dimensional orientable manifolds is 
described in \cite{FT}: the second method is equivalent to the 
introduction of a spin\( ^{c} \)-structure on \( M \). 

In the physicist's  local approach one 
introduces---following Schr\"o\-din\-ger---a spinor field as a smooth 
map \( \psi:M\to S \). The set of 
all such fields is a module \( {\mathcal S} \) over the ring \( 
{\mathcal C} \) of smooth functions on the  
Riemannian manifold \( M \) assumed here to be
 of dimension \( m=2n \).
 Let \( \nabla  \) be a covariant 
derivative in the module \( {\mathcal V} \) of vector fields: if \( 
u,v\in{\mathcal V} \), then \( \nabla_{u}v\in{\mathcal V} \) is 
the covariant derivative of \( v \) in the direction of \( u \). The 
representation \( \gamma \) mentioned above associates with a vector 
field \( u \) and a spinor field \( \psi \) another spinor field \( 
\gamma(u)\psi \); the map \( {\mathcal V}\times {\mathcal 
S}\to{\mathcal S} \), \( (u,\psi )\mapsto \gamma(u)\psi \),  is 
bilinear and \( \gamma(u)f\psi =f\gamma(u)\psi  \) for every \( 
f\in{\mathcal C} \); moreover, it has the Clifford property: 
\begin{equation}
\gamma(u)^{2}\psi =g(u,u)\psi.
\label{e:guu}
\end{equation}
 One postulates now the existence of a 
{\it spinor covariant derivative\/} \( \nabla^{\rm
s}_{u}:{\mathcal S}\to {\mathcal S} \) compatible with \( \nabla \) 
in the sense that
\begin{equation}
\nabla^{\rm s}_{u}(\gamma(v)\psi)=\gamma(\nabla_{u}v)\psi
+\gamma(u)  \nabla^{\rm s}_{u}\psi\quad\text{for}\;\; u,v\in
{\mathcal V}\;\;\text{and}\;\; \psi\in{\mathcal S}.
\label{e:comp}
\end{equation}
Let \( (e_{\mu}) \), where \( \mu=1,\dots,m \), be a field of  frames 
and let \( (e^{\mu}) \) be the dual field of coframes;
they need not be orthonormal; e.g., given local  coordinates \( (x^{\mu}) 
\), one can take \( e^{\mu}=\rd x^{\mu} \). The {\it field\/} \( 
\gamma_{\mu}=\gamma(e_{\mu}):M\to{\End} S \) satisfies
\(
\gamma_{\mu}\gamma_{\nu}+
\gamma_{\nu}\gamma_{\mu}=2g_{\mu\nu}\),
where \( g_{\mu\nu}=g(e_{\mu},e_{\nu})
\).
The coefficients of the 
linear connection defined by \( \nabla \) can be read off from
\(
\nabla_{e_{\nu}}e_{\mu}=e_{\rho}\varGamma^{\rho}_{\mu\nu}.
\)
The spinor covariant derivative of \( \psi \) in the direction of \( u 
\) can be written as the contraction
\(
\nabla^{\rm s}_{u}\psi=u\lrcorner D\psi,
\)
where \( D\psi \) has the form \eqref{e:defD}. The compatibility 
condition \eqref{e:comp} is equivalent to
\begin{equation}
\rd \gamma_{\mu}+[\omega,\gamma_{\mu}]-
\varGamma^{\rho}_{\mu\sigma}\gamma_{\rho}e^{\sigma}=0.
\label{e:SK}
\end{equation}
The metricity of \( \nabla \) can be justified by 
covariant-differentiating both 
sides of \eqref{e:guu} and using \eqref{e:comp}. 
If \( \omega \) is a solution of \eqref{e:SK} and \( A \) is a 
complex-valued 1-form 
on \( M \),  then  \( \omega+\ri A\id_{S} \) is another solution. 
Therefore, the connection \( \omega \) can be interpreted as 
including an interaction with the electromagnetic field (of potential 
equal to the real part of \( A \)).
Such an interpretation has been clearly formulated by Fock \cite{Fo} 
and Schr\"odinger \cite{Sch}: they may thus be considered 
as  precursors  of the idea of 
spin\( {}^{c} \)-structures.

\section*{Examples}
\subsection*{Hypersurfaces in ${\mathbb R}^{m+1}$}
Every hypersurface \( M \) in the Euclidean \( \text{space} \) 
\( {\mathbb R}^{m+1} \), defined by an isometric immersion 
\( f:M\to {\mathbb R}^{m+1}  \), has 
a \( \Pin_{0,m} \)-structure, canonically defined by \( f \). 
Moreover, the {\it associated bundle\/}
 \( \varSigma \) of spinors on \( M \) is {\it 
trivial\/}: it is isomorphic to the Cartesian product \( M\times S 
\) and a spinor field can be (globally!) described by a funtion \( 
\psi:M\to S \). A Dirac (resp., Cartan) spinor field on an even 
(\(\text{resp.,} \)  odd) dimensional hypersurface is the restriction of a Pauli 
(resp., Dirac) field on the surrounding space. In terms of the 
trivialization of \( \varSigma \),  the Dirac operator 
assumes a rather simple form \cite{AT1}. 
Let \( \nu_{i} \), where \( i=1,\dots,m+1 \), be 
the Cartesian components of a unit, normal vector field on \( M \). Let
\( m=2n  \) or \( 2n-1 \);
introduce the Dirac \( 2^{n}\times 2^{n} \) matrices \( \gamma_{i} \) 
satisfying
\(
\gamma_{i}\gamma_{j}+\gamma_{j}\gamma_{i}=-2\delta_{ij},\quad 
i,j=1,\dots, m+1.
\)
Then
\begin{equation}
{\mathcal D}=\tfrac{1}{2}(\gamma^{k}\nu_{k})(\gamma^{i}\gamma^{j}
(\nu_{j}\partial_{i}-\nu_{i}\partial_{j}) -\text{div}\, \nu),
\label{e:D}
\end{equation}
where \( \text{div}\,\nu \) is the intrinsic divergence of \( \nu \), 
\( 
\text{div}{\nu}=\sum_{i,j}(\delta_{ij}-\nu_{i}\nu_{j})\partial_{i}\nu_{j} \).
If \( M \) is the hyperplane of equation \( x^{m+1}=0 \), then \( 
\nu_{i}=\delta_{i}^{m+1} \) and \( 
{\mathcal D}=\sum_{\mu=1}^{m}\gamma^{\mu}\partial_{\mu} \). 
Formula \eqref{e:D} has been used to find, in a simple manner, the 
spectrum and the eigenfunctions of the 
Dirac operator on spheres \cite{AT2}.
\subsection*{Spheres, projective spaces, and quadrics}

1. The spin structures on {\it spheres\/} are 
well-known: for every \( m>1 \), 
the \( m \)-dimensional sphere \( {\mathbb S}_{m} \) has a unique spin 
structure; in the style of \eqref{e:sstr} 
it is given by the sequence of maps
\(
{\Spin}_{m}\to{\Spin}_{m+1}\to{\SO}_{m+1}\to{\mathbb S}_{m}.
\)

The spectrum of \( {\mathcal D}\) on \( {\mathbb S}_{m} \) is of the form:

\begin{picture}(400,85)(22,-60)
\put(10,0){\line(1,0){360}}
\put(55,-2.7){$\bx$}\put(33,-17){\small $-\tfrac{1}{2}m\!-\!2$}
\put(100,-2.7){$\bullet$}\put(82,-17){\small $-\tfrac{1}{2}m\!-\!1$}
\put(145,-2.7){$\bx$}\put(134,-17){\small $-\tfrac{1}{2}m$}
\put(190,-0.6){.}
\put(280,-2.7){$\bx$}\put(269,-17){\small $\tfrac{1}{2}m\!+\!1$}
\put(325,-2.7){$\bullet$}\put(310,-17){\small 
$\tfrac{1}{2}m\!+\!2\quad\cdots$}
\put(235,-2.7){$\bullet$}
\put(228,-17){\small $\tfrac{1}{2}m$}
\put(15,-17){$\cdots$}
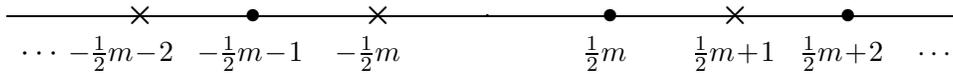
\put(35,-38){Fig. 1. The spectrum of the Dirac operator on the 
$m$-sphere.}
\end{picture}

\noindent The eigenfunctions \( \psi:{\mathbb S}_{m}\to S \) are either 
symmetric or antisymmetric; these two types of symmetries are 
indicated here by bullets and crosses; which of the eigenfunctions 
(bullets or crosses) are even depends on the trivialization of the 
bundle of spinors; only the relative parity matters.

2. The real projective spaces 
\( {\mathbb R}\text{P}_{m}={\mathbb S}_{m}/
{\mathbb Z}_{2} \) are orientable iff \( m \) is odd; for \( m>1 \) 
there are either two inequivalent  (s)pin structures or none \cite{DT}:
\begin{equation*}
\begin{array}{ccccccc}
m\quad \equiv & 0 & 1 & 2 & 3& \text{mod} & 4\\
\text{structure:}   & {\Pin}_{m,0} &
 \text{no} &{\Pin}_{0,m} & {\Spin}_{m} &&
\end{array}
\end{equation*}
Since \( {\mathbb R}\text{P}_{m}\) is locally isometric to   \( {\mathbb 
S}_{m} \), the spectrum of \( {\mathcal D}\)  
on such a \( \text{space} \)  can be obtained from 
that of the sphere: the symmetric eigenfunctions descend to one 
(s)pin structure on \( {\mathbb R}\text{P}_{m}\) and the antisymmetric 
functions---to the other; these spectra are thus asymmetric.

3. The real projective quadrics are defined as conformal 
compactifications of pseudo-Euclidean spaces; they generalize the 
Penrose construction of compactified Minkowski space-time. The quadric
\( {\mathbb S}_{k,l}=({\mathbb S}_{k}\times {\mathbb S}_{l})/{\mathbb 
Z}_{2}\) admits two natural metrics, descending from the spheres: a 
proper Riemannian metric and a pseudo-Riemannian one, of signature
\( (k,l) \). The quadrics \( {\mathbb S}_{k,0} \) and \( {\mathbb S}_{0,k} \)
can be identified with \( {\mathbb S}_{k} \);  a quadric is said to be 
{\it proper\/} if \( kl\neq 0 \). A proper quadric is orientable iff 
its dimension is even. If \( kl>1 \), then \( {\rm H}^{1}({\mathbb 
S}_{k,l},{\mathbb Z}_{2}) ={\mathbb Z}_{2}\); therefore, for
\( kl>1 \), the quadric has either 2 inequivalent (s)pin structures 
or none.  The following  table, based on \cite{CGT1},
 summarizes the results on the existence of (s)pin 
structures on \( {\mathbb S}_{k,l} \) for \( kl>1 \):
\begin{center}
\begin{tabular}{|c|c|c|c|}
\hline\hline
$k+l=2n$ & $n$ & proper Riem. & pseudo-Riem.\\
\hline
either $k$ or $l=1$ &any & yes & yes\\
 $k$ and $l$ even &  even & no & yes\\
$k$ and $l$ even &  odd &yes & no\\
 $k$ and $l$ odd &  even & no & no\\
$k$ and $l$ odd &  odd &yes & yes\\
\hline\hline
$k$ even and $l$ odd &&&\\
\hline
$k+l\equiv 1\bmod 4 $ &&${\Pin}_{0,k+l}$ &${\Pin}_{l,k}$\\
$k+l\equiv 3\bmod 4 $ &&${\Pin}_{k+l,0}$ &${\Pin}_{k,l}$\\
\hline\hline
\end{tabular}
\end{center}
For example, the quadric \( {\mathbb S}_{3,5} \) has no spin 
structure for either of the metric tensors.

The spectrum of the Dirac operator on the projective quadrics can be 
obtained from that of the spheres; this is facilitated by the 
following Lemma: if \( \lambda_{i} \) is an eigenvalue of the Dirac 
operator on a Riemannian spin manifold \( M_{i} \), \( i=1,2 \), then 
the numbers \( \sqrt{\lambda_{1}^{2}+\lambda_{2}^{2}} \) and 
\( -\sqrt{\lambda_{1}^{2}+\lambda_{2}^{2}} \) are eigenvalues of the 
Dirac operator on \( M_{1}\times M_{2} \) \cite{CGT}.
\section*{Action of space and time reflections on spinors}
\subsection*{Charge conjugation}
Space and time reflections seem to be closely related to charge 
conjugation. Consider the Dirac equation
\begin{equation*}
(\gamma^{\mu}(\partial_{\mu}-\ri qA_{\mu})-m)\psi=0
\end{equation*}
for the Dirac wave function \( \psi:{\mathbb R}^{2n}\to S \) of a 
particle of mass \( m \) and charge \( q \) moving in a \( 2n 
\)-dimensional flat space-time with a metric tensor of signature \( 
(2n-1,1) \) and an electromagnetic field with potential 
\( A_{\mu} \). Let \( C:S\to \bar{S} \) be the isomorphism such that \( 
\overline{\gamma_{\mu}}=C\gamma_{\mu}C^{-1} \), where bar denotes 
complex conjugation; the {\it charge 
conjugate\/} spinor field \( {\mathsf C}\psi=\overline{C\psi} \) 
satisfies the equation
\begin{equation*}
(\gamma^{\mu}(\partial_{\mu}+\ri qA_{\mu})-m){\mathsf C}\psi=0.
\end{equation*}
If \( \psi\sim\exp(-\ri Et) \), then \( {\mathsf C}\psi\sim\exp(+\ri Et) \):
charge conjugation is said to transform particles into antiparticles.
\subsection*{Wigner's time inversion}
Let \( \psi:{\mathbb R}^{3}\times {\mathbb R}\to {\mathbb C} \) be a 
wave function in non-relativistic quantum theory. The time-reversed 
wave function \( {\mathsf T}_{W}\psi \) is defined by \cite{W}
\begin{equation*}
({\mathsf T}_{W}\psi)({\br},t)=\overline{\psi({\br},-t)}. 
\end{equation*}
If  \( \psi \) is a solution of the 
Schr\"odinger equation with
a time-independent, real potential, then so is \( {\mathsf T}_{W}\psi \). 
If  \( \psi\sim\exp(-\ri Et) \), then 
also \( {\mathsf T}_{W}\psi\sim\exp(-\ri Et) \).
\subsection*{Time inversion of spinor fields:  Feynman {\em versus} 
Wigner}
In the relativistic theory, there are two ways of defining time 
inversion \cite{WP}. Recall first the general statement about the 
relativistic invariance of the free Dirac equation in Minkowski space.
Let \( S \) denote, as before, the complex vector space of Dirac 
spinors. A representation of the Clifford algebra \( \Cl_{3,1} \) in \( 
S \) being given in terms of the Dirac matrices \( \gamma_{\mu} \),
 one can identifiy the group \( {\Pin}_{3,1} \) with 
a subgroup of \( \GL(S) \)  and embed  \( {\mathbb R}^{4} \) in \( \End 
S \) by \( x\mapsto x^{\mu}\gamma_{\mu} \), as usual. There is the 
exact sequence
\(
1\to {\mathbb Z}_{2}\to {\Pin}_{3,1}\xrightarrow{\Ad}{\Ort}_{3,1}\to 1,
\)
where \( \Ad(U)x=UxU^{-1} \). 
There is a similar, but inequivalent,
   extension of \( 
\Ort_{3,1} \) by \( {\mathbb Z}_{2} \) corresponding to \( {\Pin}_{1,3} 
\), as well as several other extensions described in \cite{LD}; see 
also \cite{CG} and the references given there.
Every \( U\in {\Pin}_{3,1} \) acts on spinor fields by sending a 
solution   \( 
\psi:{\mathbb R}^{4}\to S \) of the free Dirac equation 
to another solution \( {\mathsf U}\psi \),
\begin{equation*}
({\mathsf U}\psi)(x)=U(\psi(\Ad(U^{-1})x)).
\end{equation*}
In particular, with  \( \gamma_{4} \) and 
\(\gamma_{1}\gamma_{2}\gamma_{3}\in {\Pin}_{3,1}  \), there are 
associated the  
space and time inversion operators  \( {\mathsf P} \)  and \( {\mathsf 
T} \), respectively. The operator \( 
{\mathsf T} \) is the {\it geometrical\/} time inversion; if  \( 
\psi\sim\exp(-\ri Et) \), then  \( {\mathsf T} \psi\sim\exp(+\ri Et) \).
 One can justify the interpretation of \( {\mathsf T} \) as the time 
inversion operator   by the Feynman idea of 
viewing antiparticles as  particles travelling backwards in 
time. Physicists favour nowadays the {\it Wigner 
time inversion\/}
\begin{equation*}
{\mathsf T}_{W}={\mathsf T}\circ {\mathsf C}.
\end{equation*}
Since charge conjugation is involutory, \( {\mathsf C}^{2}=\id \), 
the product of operators considered in the PCT theorem is equivalent 
to \( {\mathsf P}\circ {\mathsf T} \). This is the space-time 
reflection \( {\mathsf R} \) corresponding to \( \gamma_{5} \). 
The space-time reflection is in the connected component of the 
identity of the ``complex'' Lorentz group (=\( \SO_{4}({\mathbb C}) \)).
 The idea 
underlying the PCT theorem is that, in a quantum field theory invariant 
only with respect to the connected component of the Poincar\'e group,
holomorphic functions such as the 
vacuum expectation values of field operators,
are also invariant with respect to the 
``complex rotation'' \( {\mathsf R}  \)
\cite{SW}. 

In non-relativistic quantum mechanics, there is no place for \( {\mathsf 
T} \) because that theory is obtained as a limit of the relativistic 
theory implying, for a free particle, 
the removal of all negative energy states. 

\subsection*{Space and time inversion of Weyl spinors}

Complex conjugation appears also in the realization of space and time 
reflections in the space of Weyl (chiral) spinors proposed by 
Staruszkiewicz \cite{AS}; see also \cite{CG}. 

Recall that the connected component of the group \( {\Spin}_{3,1} \) 
is isomorphic to \( {\SL}_{2}({\mathbb C}) \). It has two 
inequivalent representations in the spaces of 2-component  (Weyl) 
dotted  and undotted spinors.\footnote{This terminology is due to 
van der Waerden; many people follow now 
 Penrose and use the names: primed 
and unprimed ``reduced''  spinors; see \cite{AT3}  for references and 
further comments.} 
These representations are complex 
conjugate one to another; their direct sum defines the 
representation in the space of Dirac spinors,
\begin{equation*}
a\mapsto \begin{pmatrix}a&0\\0&\bar{a}\end{pmatrix},\quad a\in
{\SL}_{2}({\mathbb C}).
\end{equation*}
This decomposable representation is a restriction to \( {\SL}_{2}({\mathbb 
C}) \) of the representation of \( {\Pin}_{3,1} \) determined by the Dirac 
matrices
\(
\gamma_{1}=\bigl(\begin{smallmatrix}0&\ri\sigma_{1}\\
-\ri\sigma_{1}&0\end{smallmatrix}\bigr),\)
\(\gamma_{2}=\bigl(\begin{smallmatrix}0&I\\I&0
\end{smallmatrix}\bigr)\),
\(\gamma_{3}=\bigl(\begin{smallmatrix}0&\ri\sigma_{3}\\
-\ri\sigma_{3}&0\end{smallmatrix}\bigr)\),
\(\gamma_{4}=\bigl(\begin{smallmatrix}0&\sigma_{2}\\
-\sigma_{2}&0\end{smallmatrix}\bigr)\).
In this representation one has \( C=\gamma_{2} \) and
 a Dirac spinor of the form
\begin{equation}
\psi=\begin{pmatrix}u\\\bar{u}\end{pmatrix},\quad \text{where}\quad 
u\in {\mathbb C}^{2},
\label{e:Ma}
\end{equation}
is real in the sense that it satisfies the Majorana condition, \( 
{\mathsf C }\psi=\psi \). Space and time inversions, as defined in 
the previous section, induce the following transformations of the 
Weyl part \( u \) of the Majorana spinor \eqref{e:Ma},
\[
{\mathsf P }: u\mapsto \sigma_{2}\bar{u}\quad\text{and}\quad
 {\mathsf T }: u\mapsto \ri \sigma_{2}\bar{u},
\]
respectively.


\begin{thebibliography}{99}

\bibitem{BR} Barut, A. and R\c aczka, R., {\it Theory of group 
representations and applications\/}, Warszawa: PWN, 1977.
\bibitem{BT} Benn, I. M., and Tucker, R. W., {\it An introduction to 
spinors and geometry with applications in physics\/}, Bristol: Hilger, 
1988.
\bibitem{Bo} Bourbaki, N., {\it Alg\`ebre\/}, 
Paris: Hermann, Masson, and Diffusion C.C.I.S.  1959--80.
\bibitem{CGT1} Cahen, M., Gutt, S. and Trautman, A., {\it J. Geom. 
Phys.}, {\bf 10}, 127--154 (1993) and {\bf 17}, 283--297 (1995).
\bibitem{CGT} Cahen, M., Gutt, S. and Trautman, A., article in: {\it Clifford 
algebras and their applications in mathematical physics\/}, V. 
Dietrich {\it et al.} (eds.), Dordrecht: \( \text{Kluwer} \), 1998, pp. 391--399.
\bibitem{CG} Chamblin, A. and Gibbons, G. W., {\it Class. Quantum 
 Grav.\/} {\bf 12}, 2243--2248 (1995).
\bibitem{LD} D\c abrowski, L., {\it Group actions on spinors}, 
Naples: Bibliopolis, 1988, pp. 8--13.
\bibitem{DT} D\c abrowski, L. and Trautman, A., {\it J. Math. 
Phys.\/} {\bf 27}, 2022--28 (1986).
\bibitem{Fo} Fock, V., {\it Z. Physik\/}, {\bf 57}, 261 (1929).
\bibitem{F} Friedrich, T., {\it Dirac-Operatoren in der Riemannschen 
Geometrie\/}, Wiesbaden: Vieweg, 1997.
\bibitem{FT} Friedrich, T. and Trautman, A., {\it Clifford structures 
and spinor bundles}, Sfb 288 Preprint No. 251, Inst. f\"ur Reine 
Mathematik, Humboldt University, Berlin 1997.
\bibitem{K} Karrer, G., {\it Ann. Acad. Fennicae\/}, Ser. A, Math. 
336/{\bf 5}, 1--16 (1963).
\bibitem{WP} Pauli, W., article  in: {\it Niels Bohr and the 
development of physics},  W. Pauli and L.~Rosenfeld (eds.),  London 
and New York: Pergamon Press, 1955, pp. 30--51.
\bibitem{Sch} Schr\"odinger, E., {\it Sitzungsber. preuss. Akad. 
Wissen.\/}, Phys.-Math.  Kl. {\bf XI}, 105--128 (1932).
\bibitem{AS} Staruszkiewicz, A., {\it Acta Physica Polonica\/} {\bf 
B7}, 557--565 (1976).
\bibitem{SW} Streater, R. F. and Wightman, A. S., {\it PCT, spin \& 
statistics, and all that\/}, New York: Benjamin, 1964.
\bibitem{AT1} Trautman, A., {\it J. Math. Phys. } {\bf 33}, 4011--4019 
(1992).
\bibitem{AT2} Trautman, A., {\it Acta Physica Polonica\/} {\bf 
B26}, 1283--1310 (1995).
\bibitem{AT3} Trautman, A., {\it Contemporary Mathematics\/} {\bf 203},
3--24 (1997).
\bibitem{W} Wigner, E. P., {\it G\"ottinger Nachrichten}, Math. Phys. 
Kl.,  pp. 546--559 (1932). 

\end{thebibliography}
\end{document}